\documentclass[useAMS,usegraphicx]{mn2e}

\newcommand{\Teff}{\mbox{$T_{\mbox{\scriptsize eff}}\,$}}
\newcommand{\Rsolar}{\mbox{R$_{\odot}$}}
\newcommand{\Msolar}{\mbox{M$_{\odot}$}}

\newcommand{\kms}{\mbox{${\rm km\,s}^{-1}$}}

\newcommand{\rxj}{\mbox{\rm RX\,J2130.6+4710}}

\title[\rxj]
{RX\,J2130.6+4710 -- an eclipsing white dwarf\,--\,M-dwarf binary star.
\thanks{Based on observations made with William Herschel, Isaac Newton and
Jacobus Kapteyn telescopes operated on the
island of La Palma by the Royal Greenwich Observatory in the Spanish
Observatorio del Roque de los Muchachos of the Instituto de Astrofisica de
Canarias.}}
\author[P.~F.~L. Maxted et~al.]
{P.~F.~L. Maxted$^{1,2}$, T.~R. Marsh$^{3,2}$, L. Morales-Rueda$^{4,2}$, 
M.~A. Barstow$^5$
  \newauthor P.~D. Dobbie$^5$, M.~R., Schreiber$^{6}$,  V.~S. Dhillon$^7$ and
C. S. Brinkworth$^2$  \\
 $^1$School of Chemistry \& Physics, Keele University, Staffordshire, ST5~5BG,
 UK. \\
 $^2$University of Southampton, Department of Physics \& Astronomy,
 Highfield, Southampton, S017 1BJ, UK. \\
 $^3$Department of Physics, University of Warwick, Coventry, CV4 7AL, UK.\\
 $^4$Department of Astrophysics, Faculty of Science, University of Nijmegen,
  P.O.~Box~9010, 6500 GL Nijmegen, The Netherlands. \\
 $^5$Department of Physics \& Astronomy, University of Leicester, University
Road, Leicester LE1 7RH, UK. \\
$^6$Universit\'{e} Louis Pasteur, Observatoire de Strasbourg, 11 rue de
l'Universit\'{e}, 67000 Strasbourg, France\\
$^7$Department of Physics and Astronomy, University of Sheffield, Sheffield,
S3~7RH \\
}

\date{Accepted --- Received ---}

\pagerange{\pageref{firstpage}--\pageref{lastpage}}

\pubyear{~}

\begin{document}

\maketitle

\label{firstpage}
\begin{abstract} 
We report the detection of eclipses in the close white-dwarf\,--\,M-dwarf
binary star \rxj.  We present lightcurves in the B, V and I bands and fast
photometry obtained with the three channel CCD photometer Ultracam of the eclipse in the $u^{\prime}$, $g^{\prime}$ and $r^{\prime}$
bands. The depth of the eclipse varies from 3.0\,magnitudes in the
$u^{\prime}$ band  to less than 0.1\,magnitudes in the I band. The times of
mid-eclipse are given by the ephemeris  BJD(Mid-eclipse)$ =
2\,452\,785.681876(2) +  0.521\,035\,625(3) E$, where figures in parentheses
denote uncertainties in the final digit. We present medium resolution
spectroscopy from which we have measured the spectroscopic orbits of the
M-dwarf and white dwarf. We estimate that the spectral type of the M-dwarf is
M3.5Ve or M4Ve, although the data on which this is based is not ideal for
spectral classification. We have compared the spectra of the white dwarf to
synthetic spectra from pure hydrogen model atmospheres to estimate that the
effective temperature of the white dwarf is $\Teff = 18\,000 \pm 1000$\,K. We
have used the width of the primary eclipse and duration of totality measured
precisely from the Ultracam $u^{\prime}$ data combined with the amplitude of
the ellipsoidal effect in the I band and the  semi-amplitudes of the
spectroscopic orbits to derive masses and radii for the M-dwarf and white
dwarf. The M-dwarf has a mass of $ 0.555 \pm 0.023 $\,\Msolar\, and a radius
of $0.534\pm 0.053$\Rsolar, which is a typical radius for stars of this mass.
The mass of the  white dwarf is $0.554 \pm 0.017 $\Msolar\, and its radius is
$0.0137\pm 0.0014$\Rsolar, which is  the  radius expected for a carbon-oxygen
white dwarf of this mass and effective temperature. The lightcurves are
affected by frequent flares from the M-dwarf and the associated dark spots on
its surface can be detected from the distortions to the lightcurves and radial
velocities. \rxj\ is a rare example of a pre-cataclysmic~variable star which
will start mass transfer at a period above the period gap for cataclysmic
variables.

 \end{abstract} 

\begin{keywords}
stars: binaries: close -- stars: white dwarfs  -- stars: fundamental
parameters -- stars: binaries: eclipsing -- stars: individual: \rxj\ 
\end{keywords}

\section{Introduction}
 \rxj\ is a soft X-ray source which was detected in a survey of the galactic
plane by the ROSAT satellite (Tr\"{u}mper 1983, Motch et~al. 1991). Motch
et~al. (1997) used optical spectroscopy of stars near the position of these
X-ray sources to identify and classify optical counterparts in the region of
the Cygnus constellation. \rxj\ was identified with a faint star which they
labelled RX\,J2130.3+4709 about 12\,arcsec\ SSE of the bright G0 star HD\,204906
(=SAO\,50947, V=8.45). The spectrum of \rxj\  shows the broad Balmer lines of
a typical DA white dwarf together with narrow emission lines in the cores in
the Balmer lines and the Ca\,II H\&K lines. Also visible are molecular bands
of TiO from the Me-type companion star which is also the source of the narrow
emission lines. They classified \rxj\ as a WD+Me close binary star. Motch
et~al. report that the radial velocities of  broad Balmer lines and emission
lines move in opposite directions with an orbital period of approximately 12
hours. The white dwarf  is too cool to contribute to the X-ray flux ($1.6
\times 10^{-2} {\rm \,cts\,s}^{-1}$) which probably arises from the Me star
companion. The distance from the optical counterpart to catalogued position of
the X-ray source is almost exactly the same as the  90\% confidence radius of
26.1\,arcsec\ in the position of the X-ray source. 

 It is likely that \rxj\ is the same source as the 2E\,2128.4+4657, a
soft X-ray source detected by the Einstein satellite which was identified with
HD\,204906 by Thompson, Shelton \& Arning (1998). The position of
2E\,2128.4+4657 given in the Einstein Observatory catalog (Harris et~al. 1994)
differs by  23\,arcsec\ from the position for the optical counterpart of \rxj\
given by Motch et~al., which is comfortably within the  positional uncertainty
of 42\,arcsec\ for the X-ray source. We note that \rxj\ appears in the
``Catalogue of cataclysmic binaries, low-mass X-ray binaries and related
objects'' (Ritter \& Kolb 1998) as ``J2130+4710''. The corresponding name in
the SIMBAD database (Wenger et~al. 2000) is RX\,J2130.6+4710, which is a more
accurate reflection of the position of the optical counterpart ($\alpha=
21^h30^m18.6^s$, $\delta = +47^{\circ}10^{\prime}08^{\prime\prime}$, J2000.0)
and is the name we use throughout this paper.

 White dwarf stars are the remnants of stars less massive than about
5\,--\,8\Msolar\ which have been through the red giant phase of their
evolution. The separation of a WD+Me binary with a period of 12 hours is
2--3\Rsolar, which is much smaller than the typical size of a red giant
($\sim$100--1000\Rsolar).  This suggests that \rxj\ is a post-common envelope
binary (PCEB). In this scenario, the expanding red giant star comes into
contact with its Roche lobe and begins to transfer mass to its companion star.
This mass transfer is highly unstable, so a common envelope forms around the
companion and the core of the red giant. The drag on the companion orbiting
inside the common envelope  leads to extensive mass loss and dramatic
shrinkage of the orbit (Iben \& Livio 1993). Common-envelope (CE) evolution
affects the evolution of many binary stars and can produce some of the most
dramatic objects known, including low-mass and high-mass X-ray binaries,
novae, supernovae and AM~CVn binaries. However, it is a poorly understood
process, so it is useful to study relatively simple PCEBs like \rxj\ to see
what they can tell us about this phenomenon. 

 \rxj\ is one of a very small sample of eclipsing white dwarfs with companions
that can be directly detected in the optical spectrum. These binaries offer a
rare opportunity to accurately measure the properties of low mass stars and
white dwarfs.  In this paper we report photometric and spectroscopic
observations of \rxj. We have used these to measure the masses and radii of
the white dwarf and the M-dwarf to an accuracy of about 5\,per\,cent. We find
that the white dwarf in \rxj\ is a typical carbon-oxygen white dwarf and that
the M-dwarf has a normal radius for its mass.

\section{Observations \& Reductions}

\subsection{Photometry}

 We first observed \rxj\ using the 1m Jacobus Kapteyn telescope (JKT) on the
Island of La Palma. We obtained 478 images with an I~band filter and 520
images with a V-band filter using a TEK charge coupled device (CCD) during the
interval 1998 Aug 5--9. The exposure times were 1.5--3\,s for the I~band
images and 1.5--5\,s for the V-band images. The image scale on the CCD was
0.33\,arcsec\ per pixel. Our intention was to confirm the orbital period of the
binary from the sinusoidal reflection effect due to the irradiation of one
side of the M-dwarf by the white dwarf. We were delighted to find that the
V-band images show an eclipse 0.6\,magnitudes deep lasting about 30\,minutes
occurring at about 0030UT 1998 Aug 10. The I~band images show an eclipse about
0.1\,magnitudes deep at the same time. This is exactly what would be expected
for the eclipse of a white dwarf by an M-dwarf with a similar luminosity in
the V-band.

 We obtained further photometry of \rxj\ with the JKT during the interval 2000
August 16--22 using a SITe CCD, which gives the same image scale as the TEK
CCD.  Images were obtained using Harris UBVRI filters. The number of images
secured with each filter and exposure times are given in
Table~\ref{UBVRITable}. The shorter exposures with the B filter were used for
observations during eclipses of \rxj.

\begin{table}
\begin{center}
\caption{Exposure times, T$_{\rm exp}$, for observations of \rxj\ obtained with the JKT. $N$ is the number of useful images obtained.
\label{UBVRITable}}
\begin{tabular}{@{}crr}
\multicolumn{1}{l}{Filter} &
\multicolumn{1}{l}{$N$} &
\multicolumn{1}{c}{T$_{\rm exp}$(s)}\\
\hline
\noalign{\smallskip}
U & 6 & 120 \\
B & 1002 & 6.5\,--\,60 \\
V & 536 & 30 \\
R & 6 & 20 \\
I & 581 & 30 \\
\hline
\noalign{\smallskip}
\end{tabular}
\end{center}
\end{table}


 Primary eclipses of \rxj\ were observed with ULTRACAM and the WHT on the
nights of 17 May 2002, 24 May 2003 and 25 May 2003. Attempts to observe the
secondary eclipse on 19 May 2002 and 13 November 2003 were compromised by poor
seeing and flaring of the M-dwarf. We have subsequently discovered that the
secondary eclipse is likely to be extremely shallow (see later) so we do not
discuss these observations further in this paper.

 ULTRACAM is an ultrafast, triple-beam CCD camera. The light is split into
three wavelength colours ($u^{\prime}$, $g^{\prime}$ and $r^{\prime}$ or
$i^{\prime}$) by two dichroic beam-splitters and then passes through a filter.
The detectors are three back-illuminated, thinned, Marconi frame-transfer
1024$\times$1024 active area CCD chips with a pixel scale of
0.3\,arcsec\,pixel$^{-1}$. ULTRACAM employs frame transfer CCDs so that the
dead time between exposures is negligible (24 milliseconds). Signals from
Global Position System (GPS) satellites were used to ensure that the
time-stamp for each image is accurate to 1\,millisecond. (for further details,
see Dhillon \& Marsh 2001).

 We reduced the data using normal extraction from fixed apertures and also
Naylor's (1998) optimal extraction from apertures varied in step with the
seeing. In the latter case we used aperture radii 1.5 times the full-width at
half maximum of the image of \rxj. \rxj\ is located only 12\,arcsec away from
a much brighter G0 star so the photometry can be badly affected in poor
seeing. To estimate and correct for the extent of this we placed an aperture
on the sky at the same distance from the bright star as \rxj\, and
symmetrically located with respect to the diffraction spikes from this star. 

The bias level in the JKT images was determined from the overscan regions and
was subtracted from the image before further processing. Images of the
twilight sky devoid of any bright stars were used to determine flat-field
corrections by forming the median image of 3--5 twilight sky images in each
filter, one for each night's data. We used optimal photometry (Naylor 1998) to
determine instrumental magnitudes of the stars in each frame. We checked for
variability in the stars other than the target star in each frame before
calculating differential magnitudes between \rxj\ and a comparison star
located at  $\alpha= 21^h30^m19.7^s$, $\delta =
+47^{\circ}10^{\prime}26^{\prime\prime}$ (J2000.0).  The mean differential
magnitudes in the sense (\rxj\ $-$ comparison) in the B, V and I bands when
\rxj\ is not in eclipse are $-0.74$, 0.29 and 0.21, respectively. 

 We used an aperture with a radius of 5 pixels to measure the instrumental
magnitude of \rxj\ and the 5 standard stars in the field of PG\,1633+099 with
UBVRI magnitudes given by  Landolt (1992) . We used these to calculate the
apparent magnitudes for \rxj\ given in Table~\ref{MagTable}. The calibration
of our photometry is approximate because the proximity of HD\,204906 to \rxj\
makes it difficult to measure reliable fluxes. For this reason we only quote
one decimal place for our apparent magnitude values. Also given in
Table~\ref{MagTable} are the infrared apparent magnitudes for \rxj\ taken from
Hoard et~al. (2002).

\begin{table}
\begin{center}
\caption{Apparent magnitudes of \rxj.\label{MagTable}}
\begin{tabular}{@{}rrrrrrrr}
\multicolumn{1}{c}{U} &
\multicolumn{1}{c}{B} &
\multicolumn{1}{c}{V} &
\multicolumn{1}{c}{R} &
\multicolumn{1}{c}{I} &
\multicolumn{1}{c}{J} &
\multicolumn{1}{c}{H} &
\multicolumn{1}{c}{K$_s$} \\
\hline
\noalign{\smallskip}
14.8 & 14.9 & 14.3 & 13.70 & 12.2 &  11.21 &     10.58  &   10.36\\
\hline
\noalign{\smallskip}
\end{tabular}
\end{center}
\end{table}

\subsection{Spectroscopy}

 We observed \rxj\ with the 2.5m Isaac Newton Telescope (INT), also on the
island of La Palma, using the IDS spectrograph simultaneously with our JKT
observations in 2000 August. Most of our observations were obtained using a
narrow slit (0.9--1\,arcsec) and a 1200 line\,mm$^{-1}$ grating on the 235mm
camera with an EEV CCD. The spectra cover the wavelength region 3850--5500\AA\
at a  resolution of about 1.2\AA\ and the dispersion is about 0.48\AA\ per
pixel. A typical spectrum is shown in Fig.~\ref{SpTyFig}. The exposure times
were 1200s and we obtained arc lamp spectra approximately once per hour to
establish the wavelength scale and monitor the spectrograph drift. We also
observed one eclipse of \rxj\ with the same instrument but with a
300\,line\,mm$^{-1}$ grating and a 4\,arcsec\ wide slit orientated to include
the same comparison star used for the JKT photometry. The exposure times were
10\,seconds and there was a 4 second delay between exposures to read out the
CCD. The spectra cover the wavelength range 4215\,--\,6795\,\AA\ at a
dispersion of 3.7\AA\ per pixel.


 Extraction of the narrow-slit spectra from the images was performed
automatically using optimal extraction to maximize the signal-to-noise of the
resulting spectra (Marsh 1989). The arcs associated with each stellar spectrum
were extracted using the same weighting determined for the stellar image to
avoid possible systematic errors due to the tilt of the spectra on the
detector.  The wavelength scale was determined from a fourth-order polynomial
fit to measured arc line positions. The standard deviation of the fit to the
18 arc lines was typically 1/20 of a pixel. The wavelength scale for an
individual spectrum was determined by interpolation to the time of
mid-exposure from the fits to arcs taken before and after the spectrum to
account for the small amount of drift in the wavelength scale ($<0.1$\AA) due
to flexure of the instrument between arc spectra. We used a spectrum of
BD$+33\,2642$ obtained during the same observing run and the tabulated fluxes
of Oke (1990) to determine the calibration of counts-to-flux as a function of
wavelength in our spectra. No correction for slit losses was attempted but we
did correct the spectra for atmospheric extinction.  Statistical errors on
every data point calculated from photon statistics are rigorously propagated
through every stage of the data reduction.

To  extract the spectrophotometry taken with a  wide slit we summed 
the measured counts over 6 spatial pixels centered on the spectrum 
at every wavelength step for both \rxj\ and the comparison star after
subtracting the sky contribution estimated from a linear fit to the regions
either side of the spectra.  The  wavelength calibration was determined from a
single arc spectrum taken at the start of the night and is accurate to about
1\AA.

\subsection{Timing of INT and JKT data}

 We warn the reader here that we have not been able to establish whether the
time-stamps for the data obtained with the INT and JKT are reliable. It
is probable that the time-stamps for these data have an accuracy of better
than 1\,second. However, the observatory staff have noted that time-stamps on
data such as these can be in error by a few seconds.

\section{Analysis}

\subsection{The ephemeris}\label{EphemSect}
\begin{figure}
\includegraphics[width=0.45\textwidth]{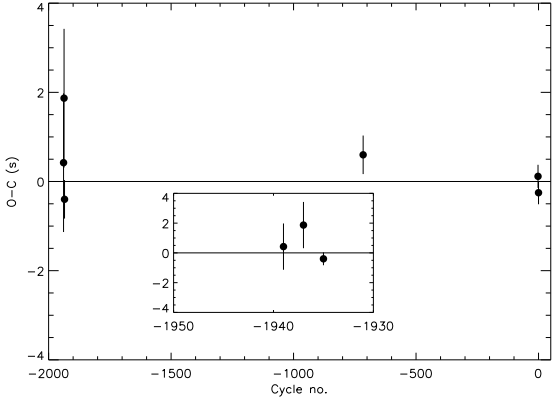}
\caption{\label{EphemFig}Residuals from a linear ephemeris for our measured
times of mid-eclipse.
}
\end{figure}

 We measured the time of mid-eclipse from our photometry using a least-squares
fit of a very simple lightcurve model. The model is based on the eclipse of
one circular disc with uniform surface brightness by another similar disc. The
parameters of the model are the radii of the discs relative to their maximum
separation, $r_1$ and $r_2$; the inclination of the orbit, $i$; the orbital
period $P$ and time of primary eclipse $T_0$; the luminosity ratio,
$l_{\lambda}$; the observed intensity outside of eclipse, $I_0$ and the mass
ratio $q$. We also include a parameter to allow for a linear variation of the
observed intensity with time. As we were only interested in the time of
mid-eclipse, we fixed the parameters $q=1$, $i=90^{\circ}$, $P=0.52104$\,d
and found the optimum values of the other parameters using the
Levenberg-Marquardt method (Press et~al. 1992) We fitted data covering each
eclipse we had observed separately.  The quality of the fits judged by-eye and
from the value of $\chi^2$ was very good (Fig.~\ref{uLCFig}). The results are
shown in Table~\ref{TminTable}.  Corrections from UTC to Barycentric Julian
date (BJD) were calculated using routines from {\sc slalib} version 2.4-5
(Wallace 2000) The times of mid-eclipse labelled ``JKT'' are from B-band
photometry taken with the JKT with uncertainties derived from the covariance
matrix of the least-squares fit. The lightcurve for the time of mid-eclipse
labelled ``INT'' was produced by summing the total counts over the wavelength
4215\,--\,6796\AA\ in our low resolution spectra of \rxj\ and the comparison
then taking the ratio of these time-series. The uncertainty was also taken
from the covariance matrix of the least-squares fit. The times of minimum
labelled ``Ultracam'' are the weighted averages of the values derived by
fitting the $u^{\prime}$, $g^{\prime}$ and $i^{\prime}$ lightcurves observed
with  Ultracam. There is an obvious flare during the eclipse observed on
2002~May~17 so we have excluded data in the range JD= 2452412.632 to
2452412.635 from the least-square fit. The uncertainties for these values are
taken from the standard deviation of the three measurements. The measurements
from the three lightcurves are consistent to the individual uncertainties
taken from the covariance matrix of the least-squares fit. We re-iterate our
warning here that timing of the data taken with the INT and JKT presented here
may be in error by a few seconds, so {\it these data should not be used to
study the long-term period changes.}

\begin{table}
\caption{Times of mid-eclipse for \rxj. \label{TminTable}}
\begin{tabular}{@{}rrl}
\multicolumn{1}{l}{BJD(mid-eclipse)} &\multicolumn{1}{l}{ Cycle } & Source \\
\hline
\noalign{\smallskip}
2451775.393803  $\pm$ 0.000018  &$-1939$ & JKT$^a$ \\
2451776.435891  $\pm$ 0.000018  &$-1937$ & JKT$^a$ \\
2451777.477936  $\pm$ 0.000005  &$-1935$ & INT$^a$ \\
2452412.620375  $\pm$ 0.000005  &$ -716$ & Ultracam \\
2452784.639806  $\pm$ 0.000003  &$   -2$ & Ultracam \\
2452785.681873  $\pm$ 0.000003  &$    0$ & Ultracam \\
\hline
\noalign{\smallskip}
\multicolumn{3}{l}{$^a$These data are not suitable for long-term period studies.}
\end{tabular}
\end{table}

 A least-squares fit to these data gives the following linear ephemeris:
\begin{eqnarray*}
 {\rm BJD (Mid-eclipse)} &= (2\,452\,785.681876 \pm 0.000\,002) \\
                  &+  (0.521\,035\,625 \pm 0.000\,000\,003) E. 
\end{eqnarray*}

 All phases quoted in this paper have been calculated with this ephemeris. 


\subsection{Spectral Type of the M-dwarf}
\begin{figure*}
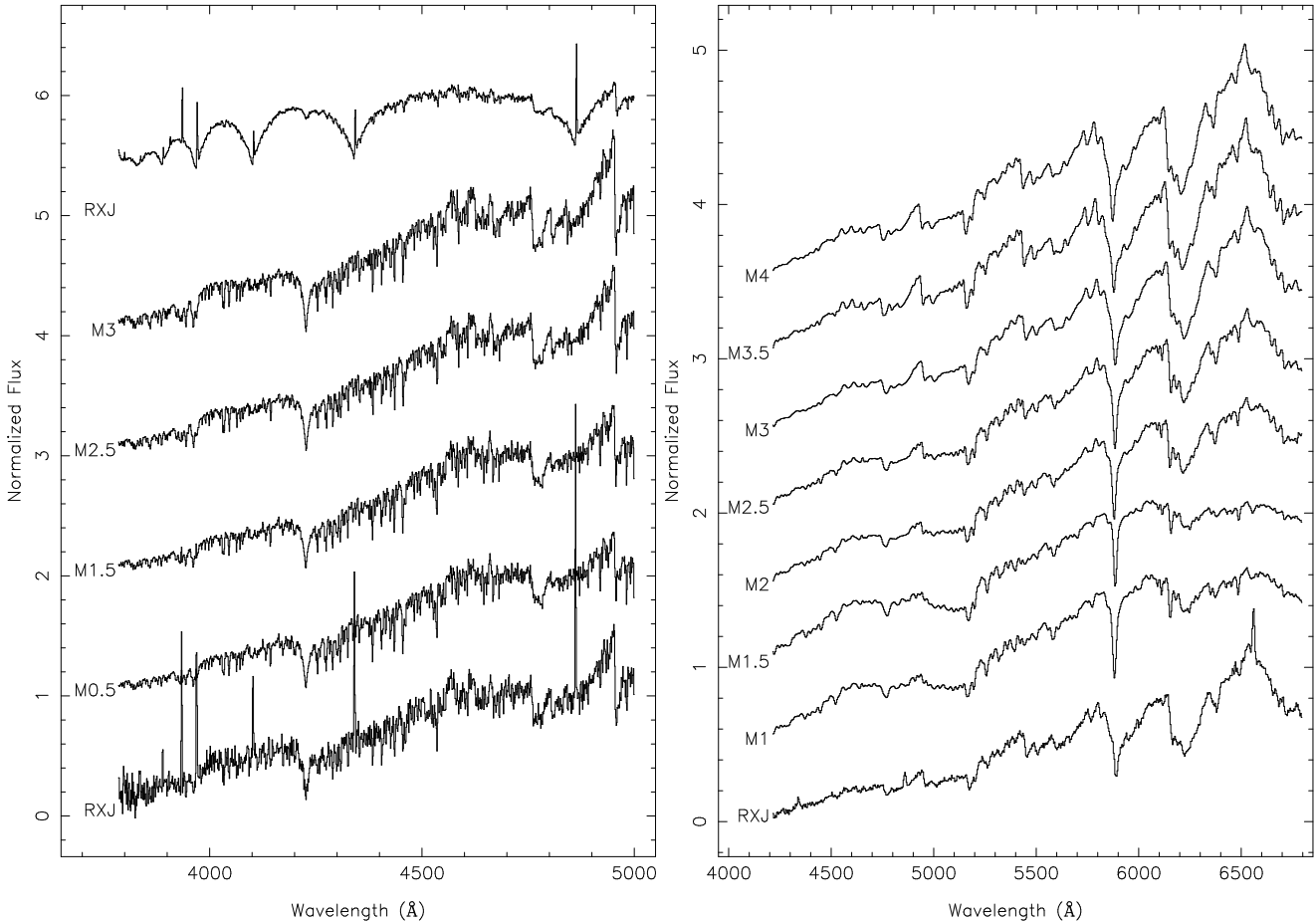

\includegraphics[width=0.49\textwidth]{SpTyBlue.ps}
\includegraphics[width=0.49\textwidth]{SpTyRed.ps}
\caption{{\bf Left panel} From bottom-to-top: spectrum of \rxj\  taken at
mid-eclipse, LFT 1580 (M0.5), LFT 1751 (M1.5), LFT 1466 (M2.5), LFT 93 (M3),
mean of 4 spectra of \rxj\  taken near phase 0.75. The spectra have been
normalized  at 5100\AA\ and offset by 1 for clarity. {\bf Right panel}  From
bottom-to-top:  mean spectrum of \rxj\  taken at mid-eclipse, LFT\,1278
(M1), LFT\,1436 (M1.5), LFT\,1580 (M2), LFT\,1751 (M2.5), LFT\,1466 (M3),
LFT\,98 (M3.5), LFT\,1431 (M4). \label{SpTyFig}}
\end{figure*}

 One of our blue narrow-slit spectra was taken during a total eclipse of \rxj\
so it only contains light from the M-dwarf. This spectrum is compared to
spectra of K- and M-dwarfs taken with the same instrument and setup in
Figure~\ref{SpTyFig}. Also shown in Figure~\ref{SpTyFig} is the average of our
wide-slit spectra taken during eclipse compared to spectra of M-dwarfs
observed with the same instrument and setup. Spectral types for the M-dwarfs
were taken from Reid, Hawley \& Gizis (1995) or Hawley, Gizis \& Reed (1996).
We estimate that the spectral type of the M-dwarf in \rxj\ is M3.5 or M4,
based principally on the strength of the TiO bands at  5448\AA\, 5847\AA\ and
6158\AA\ and the CaOH band at 5500-60\AA (Jaschek \& Jaschek 1987). The
spectra we have used for classification do not cover the TiO bands further to
the red which are more commonly used for spectral classification and we have
used only a few stars which are not standards for the comparison so the
uncertainty in the classification is likely to be at least one sub-type.

\subsection{Spectroscopic orbit of the M-dwarf\label{OrbitSect}}
 We used cross-correlation to measure the radial velocity of the M-dwarf from
the INT spectra taken in August 2000. We first removed the contribution of the
broad wings of the H$\beta$ and H$\gamma$ lines from the white dwarf by using
two Gaussian profiles to model each line. The sum of these broad Gaussian profiles and a low
order polynomial was fit to each spectrum by least squares and then subtracted 
from the spectrum prior to performing the cross-correlation. 

\begin{table*}
\caption{\label{RVMFITTable} Results of least-squares fits of a sine wave
plus a distortion to the measured radial velocities of the M-dwarf in \rxj. }
\begin{tabular}{@{}lrrrrrrrr}
\noalign{\smallskip}
Template &
\multicolumn{1}{c}{$V_{\rm temp}$} &
\multicolumn{1}{c}{ $\gamma_{\rm M}$} & 
\multicolumn{1}{c}{ $K_{\rm M}$ } & 
\multicolumn{1}{c}{A} &
\multicolumn{1}{c}{$\phi_0$} &
\multicolumn{1}{c}{$\rho$}  & 
\multicolumn{1}{c}{ $\sigma_{\rm sys}$}  &
\multicolumn{1}{c}{ $\sigma$} \\
(Sp. type) &
\multicolumn{1}{c}{(\kms)} &
\multicolumn{1}{c}{(\kms)} &
\multicolumn{1}{c}{(\kms) }&
\multicolumn{1}{c}{(\kms) }&
& & 
\multicolumn{1}{c}{(\kms) }&
\multicolumn{1}{c}{(\kms) }\\
\noalign{\smallskip}
\hline
\noalign{\smallskip}


LFT\,1580    &$-17.161$&  15.6  & 135.7  & 2.9 & 0.18 & 0.42 & 3.4 & 4.5 \\
\multicolumn{1}{c}{(M0.5)}&&$\pm$0.7&$\pm$0.7&$\pm$ 0.8 &$\pm$0.04&$\pm$0.07\\ 
\noalign{\smallskip}

LFT\,1751    &$-27.317$&  15.5  & 136.4  & 2.9 & 0.20 & 0.40 & 3.8 & 4.7 \\
\multicolumn{1}{c}{(M1.5)}&&$\pm$0.7&$\pm$0.7&$\pm$ 0.9 &$\pm$0.03&$\pm$0.06\\ 
\noalign{\smallskip}

LFT\,1466    &$ 35.880$& 16.1  & 136.7  & 2.8 & 0.23 & 0.37 & 3.9 & 4.8 \\
\multicolumn{1}{c}{(M2.5)}&&$\pm$0.7&$\pm$0.7&$\pm$ 0.9 &$\pm$0.04&$\pm$0.07\\ 
\noalign{\smallskip}

LFT\,93      &$  1.500$& 13.9  & 136.7  & 2.8 & 0.21 & 0.38 & 4.0 & 4.8 \\
\multicolumn{1}{c}{(M3)}&&$\pm$0.7&$\pm$0.7&$\pm$ 0.9 &$\pm$0.04&$\pm$0.06\\ 
\noalign{\smallskip}
\hline
\end{tabular}
\end{table*}

We used spectra of four different M-dwarfs taken with the same
instrumental setup as templates for the cross-correlation after normalization
using a low-order polynomial fit by least-squares. The
spectra were re-binned onto a uniform wavelength grid of 1835 pixels of
32\kms\ each centered on 4600\AA. The cross-correlation function was
calculated excluding 10\AA\ around the two Balmer lines, which are affected by
the emission lines from the cool star and the core of the spectral lines from
the white dwarf. The radial velocity and its uncertainty were measured from the
cross-correlation function using the method of Schneider \& Young (1980)
which, briefly, determines  the centroid of the CCF weighted in this case by a
Gaussian function with a full-width at half-maximum (FWHM) of 200\kms.

 We used a least-squares fit to the measured radial velocities, $V_r$, of the
function $V_r = \gamma_{\rm M} + K_{\rm M} \sin( 2\pi[T-T_0]/P)$  to measure
the semi-amplitude of the spectroscopic orbit for the M-star, $K_{\rm M}$. The
values of $T_0$ and the orbital period, $P$, were fixed using the ephemeris
described in section \ref{EphemSect}. From inspection of the residuals from
these fits it was clear that there was some distortion to the radial
velocities with an amplitude of a few \kms. The residuals tended to be
positive over the phases $[T-T_0]/P=$0.2\,--\,0.4 and negative over the phases
0.4\,--\,0.6. We suspect that this distortion is due to a dark spot on the
M-dwarf. We do not believe this distortion is a direct consequence of the
irradiation of the M-dwarf by the white dwarf because such a distortion would
be symmetrical around phase 0.5. To model this distortion we add the term
$A\sin([\phi - \phi_0]/\rho)$ to data in the phase range $\phi_0 < \phi <
\phi_0 + \rho$, where $\phi=2\pi[T-T_0]/P$. There may be a similar distortion
near phase 0.9, but the data near this phase range is rather sparse so we have
decided not to use a more complicated function to fit these data. The
physical origin of this distortion is a dark spot crossing the observer's
meridian on the visible disc of the M-dwarf at phase $\phi_{\rm mid}$. The
velocity of the spot will vary as $\sin(\phi -\phi_{\rm mid})$ and, since the
inclination of the system is approximately $90^{\circ}$ its visibility will
vary approximately as  $\cos (\phi -\phi_{\rm mid})$. The combination of these
effects yields a distortion of the form $\sin(\phi -\phi_{\rm mid})\cos(\phi
-\phi_{\rm mid}) = \sin [2(\phi -\phi_{\rm mid})]$. The effect of limb darkening
will be to reduce the timescale of the variation whereas the finite size of
the spot tends to increase the timescale of the distortion, so that the
velocity distortion is  approximately $ B\sin [ ( \phi
-\phi_{\rm mid})/\rho]$ with $\rho \approx 0.5$. Note that $\phi_{\rm mid} =
\phi_0+\rho/2$ so $A=-B$.

 The value of $\chi^2$ for these least-squares fits was still rather large.
We therefore added an ``external error'', $\sigma_{\rm sys}$, in quadrature to
the standard errors of the radial velocity measurements to account for
additional sources of uncertainty. These might be instrumental, e.g., motion
of the star in the slit, or intrinsic to the M-dwarf, e.g.,  weak emission
lines, or spectral features associated with chromospheric activity such as
flares or dark spots in the photosphere.  The value of $\sigma_{\rm sys}$ was
chosen to obtain a reduced $\chi^2$ value of 1.

\begin{figure}
\includegraphics[angle=270,width=0.49\textwidth]{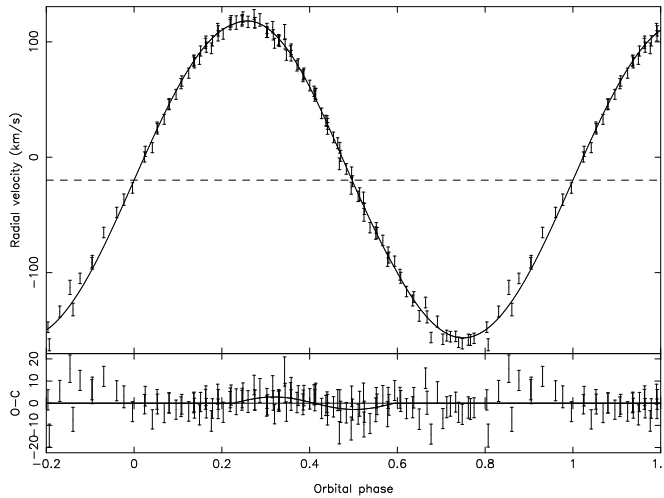}
\caption{\label{RVMFig}Radial velocities for the M-star in  RXJ\,2130.3+4709 measured by 
 cross-correlation using the spectrum of LFT\,1536. The upper panel shows the
least-squares fit to the velocities described in the text (sine wave +
distortion). The lower panel shows the residuals from the sine wave fit and
the function fit by least-squares which is used to model the distortion.}
\end{figure}

 The results of the least-squares fit of the sine wave + distortion are given
in Table~\ref{RVMFITTable}. Results are given for each of the templates used
for cross-correlation as listed in column 1. The value of $\gamma_{\rm M}$ has
been corrected for the measured radial velocity of the template, $V_{\rm
temp}$, taken from Nidever et~al. (2002).  The spectral type of the template
taken from Reid, Hawley \& Gizis (1995) or Hawley, Gizis \& Reed (1996) is
also given. The standard deviation of the residuals, $\sigma$ is given in the
final column.  An example of the measured radial velocities and the function
fit by least-squares is shown in Fig.~\ref{RVMFig}.

 We adopt the weighted mean values from Table~\ref{RVMFITTable} of
$\gamma_{\rm M} = 15.3 \pm 0.7$\kms\ and $K_{\rm M} = 136.4 \pm 0.8$\kms. The
uncertainties are the standard deviations of the four measurements with a
small additional uncertainty in $K_{\rm M}$ to allow for the partial
correlation of $K_{\rm M}$ with $A$.  These values are both about 1.5\kms\
lower than that obtained by fitting a simple sine wave to the data.

\subsection{Spectroscopic orbit of the white dwarf}

  The measurement of the radial velocities for the white dwarf in \rxj\ was
difficult because there are only three spectral lines available to us which
show a sufficiently sharp core to give reliable measurements. These spectral
regions all show many sharp absorption lines and a strong, variable emission
line from the M-dwarf. Fortunately, we have a spectrum of the M-dwarf alone
obtained during an eclipse of the white dwarf. We are able to remove much of
the contribution of the M-dwarf by subtracting this spectrum from the others.
To do this, we first shift the wavelength scale of all the spectra to remove
the orbital motion of the M-dwarf according to the orbit derived in section
\ref{OrbitSect}.  The spectra were normalized by a constant value determined
from the mean flux in the line wings. We interpolated the spectra onto three
wavelength grids centered on each of the Balmer lines of 120 pixels of 32\kms\
each. We then formed the average  of spectra in groups taken at the same
orbital phase to within 0.02 phase units.  

 We then subtracted from these 47 phase-binned spectra the spectrum obtained
in eclipse scaled by some factor so as to optimize the removal of the sharp
absorption features. This was done independently for each Balmer line as the
optimum factor varies strongly with wavelength. For the H$\delta$ line we
found that the spectrum obtained in eclipse was too noisy, so we used the
combined spectrum of the M3 and M3.5 dwarfs  LFT\,1431 and LFT\,1432 instead,
which looked very similar but was much less noisy.
 The resulting spectra show the a strong absorption line near the line centre
and the broad lines of the white dwarf with a weak core whose position varies
sinusoidally with phase. In addition, some spectra show weak emission lines
from the M-dwarf away from the line centre (Fig.~\ref{TrailHbetaFig}) and the
stronger central emission line from the M-dwarf which is not entirely removed
by the subtraction process. 

 To measure the radial velocity of the white dwarf from these spectra, we
created a model line profile for the white dwarf from the sum of three
Gaussian profiles with the same mean but independent widths and depths. We
then used  least-squares fitting to the phase-binned spectra to determine the
widths and depths of these Gaussians. Other parameters in the fit were the
coefficients of a linear polynomial used to model the continuum and the width,
height and position of a Gaussian used to model the emission line from the
M-dwarf. Finding the optimum parameters for the Gaussians and the best scaling
factor for the M-dwarf took much experimentation, but the quality of the fits
achieved judged by the $\chi^2$ value and by-eye is good. For the measured
radial velocities reported here, we fixed the widths and depths of the
Gaussians used to model the white dwarf spectrum to be the same for all the
spectra. We then repeated least-squares fit to find the radial velocity of the
white dwarf and the other free parameters of the fit. We also corrected for
the shift applied to the spectra to correct them for the orbital motion of the
M-dwarf. The phase-binned spectra around the H$\beta$ line, the model spectra
and the residuals from the fit are shown in Fig.~\ref{TrailHbetaFig}.  Two
phase-binned spectra of H$\beta$ observed near quadrature and the model
spectra are also plotted in Fig.~\ref{FitHbetaFig}.

 We used a least-squares fit to the measured radial velocities, $V_r$, of the
function $V_r = \gamma_{\rm WD} - K_{\rm WD} \sin( 2\pi[T-T_0]/P)$  to measure
the semi-amplitude of the spectroscopic orbit for the white dwarf, $K_{\rm
WD}$. The values of $T_0$ and the orbital period, $P$, were fixed using the
ephemeris derived in section \ref{EphemSect}.  Note that  $\gamma_{\rm WD} \ne
\gamma_{\rm M}$ for three reasons: {\it i.}~the gravitational redshift of
light from the white dwarf, $z = 0.635 (M/\Msolar)/(R/\Rsolar)\kms \approx 25
\kms$; {\it ii.}~the Stark effect is asymmetrical for the higher Balmer lines
resulting in pressure shifts (Grabowski, Halenka \& Madej 1987); {\it iii.}~a
tilt in the continuum will result in a systematic error in the value of
$\gamma_{\rm WD}$ measured from such broad lines.  The resulting fit for the
$H\beta$ line is shown in Fig.~\ref{RVHbetaFig}. Also shown are the radial
velocities measured for the emission line, which are a by-product of the
analysis. The values of $\gamma_{\rm WD}$ and $K_{\rm WD}$ measured from each
Balmer line are given in Table~\ref{WDRVTable}. The agreement between the
values of $K_{\rm WD}$ measured from the three lines is remarkably good, but
we suspect this is a fluke. These values are certainly affected by systematic
errors due to inaccurate subtraction of the M-dwarf spectrum, variable tilts
in the spectrum and weak emission lines from the M-dwarf. Therefore, we should
not take the error in the weighted mean as a measure of the uncertainty.
Instead we use the standard deviation of the values and adopt the value
$K_{\rm WD} = 136.5 \pm 3.8 $\,\kms.

\begin{table}
\caption{\label{WDRVTable} Results of least-squares fits of a sine wave
to the measured radial velocities of the white dwarf in \rxj. }
\begin{center}
\begin{tabular}{@{}lrrrr}
\noalign{\smallskip}
Spectral &
\multicolumn{1}{c}{ $\gamma_{\rm WD}$} & 
\multicolumn{1}{c}{ $K_{\rm WD}$ } & 
\multicolumn{1}{c}{ $\chi^2$}  &
\multicolumn{1}{c}{ $\sigma$}  \\
~ line &
\multicolumn{1}{c}{(\kms)} &
\multicolumn{1}{c}{(\kms)} &
\multicolumn{1}{c}{(n=47) }&
\multicolumn{1}{c}{(\kms) } \\
\noalign{\smallskip}
\hline
\noalign{\smallskip}
H$\beta$ & 39.2$\pm$2.6 & $139.2\pm3.4$ & 43.0 & 17.9  \\
H$\gamma$ & 58.4$\pm$2.7 & $133.9\pm3.8$ & 43.1 & 21.4 \\
H$\delta$ & 51.7$\pm$2.9 & $135.5\pm4.1$ & 52.7 & 22.3  \\
\noalign{\smallskip}
\hline
\end{tabular}
\end{center}
\end{table}

\begin{figure*}
\includegraphics[width=0.3\textwidth]{TrailHbetaFigSpec.ps}
\includegraphics[width=0.3\textwidth]{TrailHbetaFigFit.ps}
\includegraphics[width=0.3\textwidth]{TrailHbetaFigRes.ps}
\caption{Trailed grey-scale spectra of the H$\beta$ line in \rxj.
Left-to-right: observed, phase binned spectra after subtraction of the M-dwarf
spectrum; least-squares fit; residuals from the fit. For the spectra, white is
an intensity value of 0.5, black is 1.0. For the residuals the intensity scale
is $-0.1$ to $+0.1$ from white to black. \label{TrailHbetaFig} }
\end{figure*}

\begin{figure}
\includegraphics[angle=270,width=0.49\textwidth]{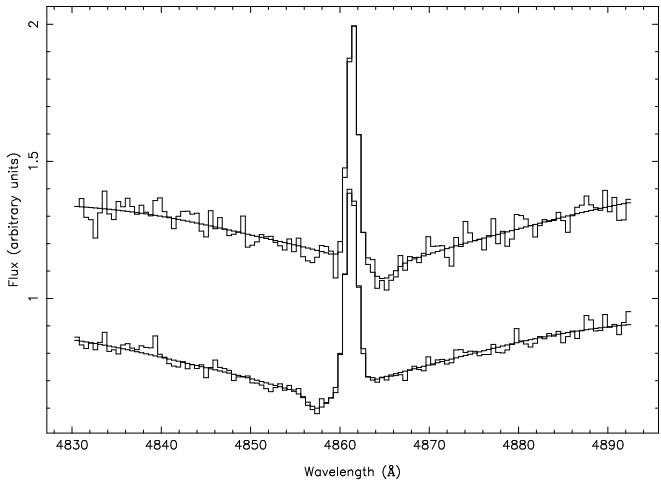}
\caption{Examples of multiple Gaussian  least-squares fits to the H$\beta$ line.
Two spectra of \rxj\ which have had the M-star spectrum subtracted offset by
0.4 units are shown (thin line) together with the model spectra (thick line).
\label{FitHbetaFig}}
\end{figure}

\begin{figure}
\includegraphics[width=0.49\textwidth]{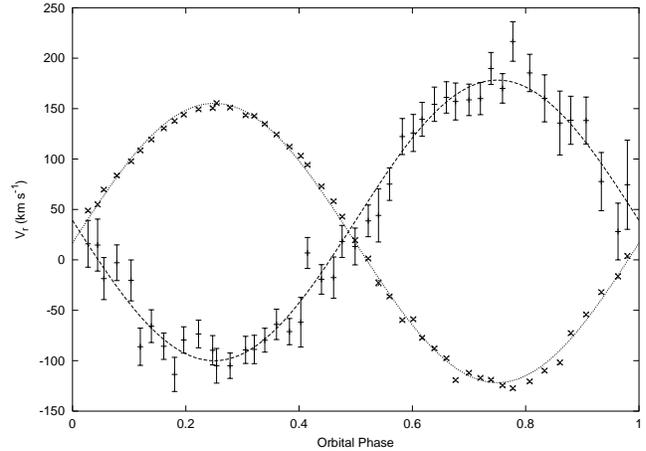}
\caption{Radial velocities of the white dwarf (+) and the emission
line from the M-dwarf ($\times$) measured from the H$\beta$ line. Sinusoidal
least-square fits are shown as solid lines. \label{RVHbetaFig}} 
\end{figure}

\subsection{The combined spectroscopic orbit}

 From the adopted value of $P$, $K_{\rm M}$ and $K_{\rm WD}$ we use Kepler's
Law to derive the minimum masses, $m_{\rm M}\sin^3i$ and $m_{\rm WD}\sin^3i$,
the mass ratio $q = m_{\rm M}/m_{\rm WD}$, and projected semi-major axis $a
\sin i$, where $m_{\rm M}$ is the mass of the M-dwarf, $m_{\rm WD}$ is the
mass of the white dwarf and $i$ is the orbital inclination. The values derived
and their uncertainties are given in Table~\ref{SpecOrbitTable}. We have taken
the value of $\gamma_{\rm WD}$ derived from the H$\beta$ line as we believe
this to be the least affected but systematic errors.

\begin{table}
\caption{\label{SpecOrbitTable} The combined spectroscopic orbit of \rxj.}
\begin{center}
\begin{tabular}{@{}lrr}
\noalign{\smallskip}
Parameter & \multicolumn{1}{l}{Value} & \multicolumn{1}{l}{Notes} \\
\noalign{\smallskip}
\hline
$T_0$&BJD  2\,452\,785.681876 & See section \ref{EphemSect} \\
$P$   & 0.521\,035\,625 d & See section \ref{EphemSect} \\
$\gamma_{\rm WD}$ & 40 \kms & Uncertain\\
$K_{\rm WD}$ &  $136.5 \pm 3.8$\kms & \\
$\gamma_{\rm M}$ &$ 15.3 \pm 0.7$\kms & \\
$K_{\rm M}$ &  $136.4 \pm 0.8$\kms & \\
$q = m_{\rm M}/m_{\rm WD}$ & $ 1.00 \pm 0.03$ & \\
$m_{\rm WD}\sin^3i$ & $ 0.548 \pm 0.016$ \Msolar & \\
$m_{\rm M}\sin^3i$  & $ 0.549  \pm 0.022 $ \Msolar & \\
 $a \sin i$ & $ 2.81  \pm 0.04 $ \Rsolar & \\ 

\noalign{\smallskip}
\hline
\end{tabular}
\end{center}
\end{table}

\subsection{The lightcurve}
 The lightcurve of \rxj\ shows that the white dwarf is eclipsed for
27\,m\,30\,s and that the ingress and egress phases of the eclipse last for
1\,m\,50\,s. The depth of the eclipse varies from 3.0\,magnitudes in the
$u^{\prime}$ band to less than 0.1\,magnitudes in the I
band (Fig.~\ref{BVILCFig}). There is no secondary eclipse apparent in our data,
which is unsurprising given that it is expected to be less than
1\,milli-magnitude deep. It is also noticeable that the amplitude of the
reflection effect is rather small, i.e., the lightcurves are almost flat
outside of the eclipse.

\begin{figure}
\includegraphics[width=0.49\textwidth]{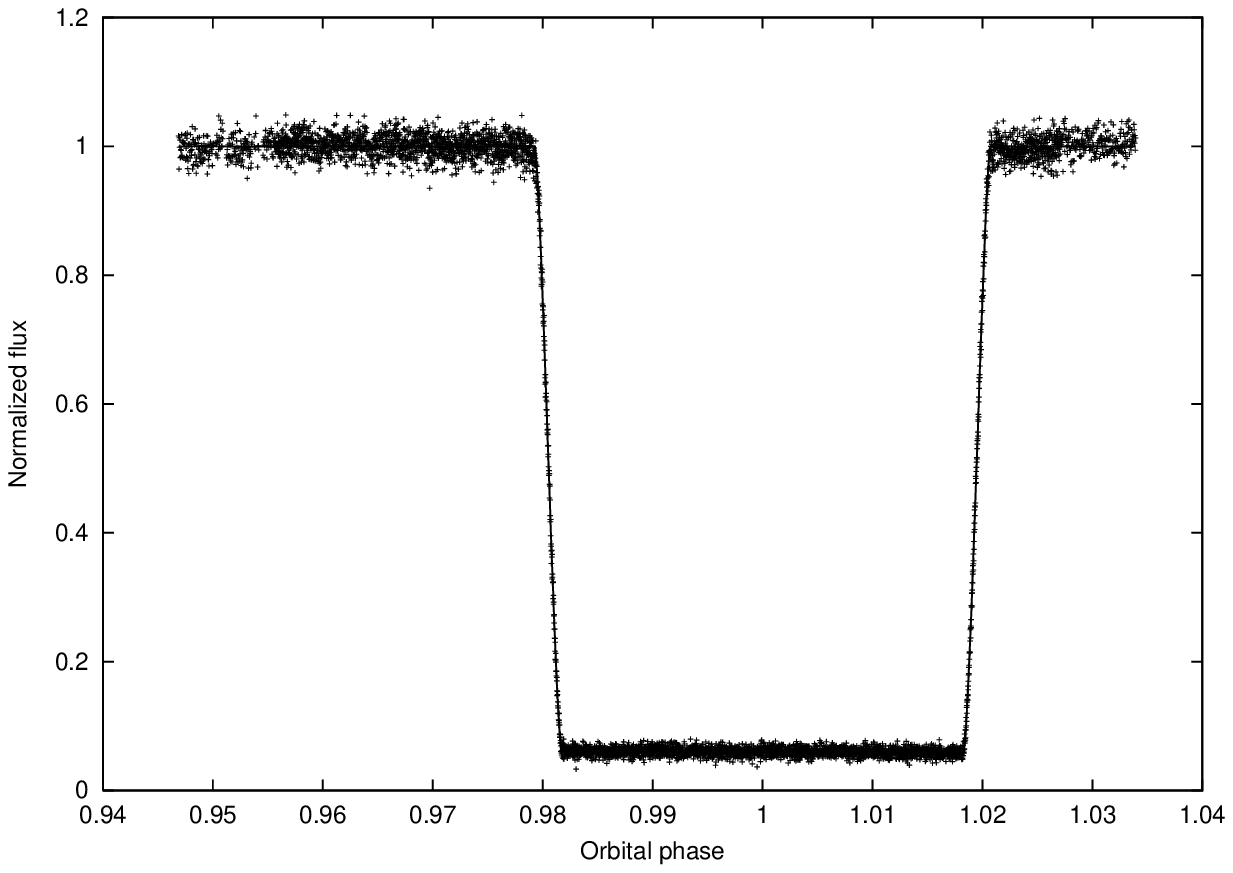}
\caption{The lightcurve of \rxj\ in the $u^{\prime}$ band (+) and
the least-squares fit for the eclipse of uniform circular discs assuming
$i=80.5^{\circ}$ (solid line, barely visible). \label{uLCFig}}
\end{figure}

\begin{figure*}
\includegraphics[width=0.99\textwidth]{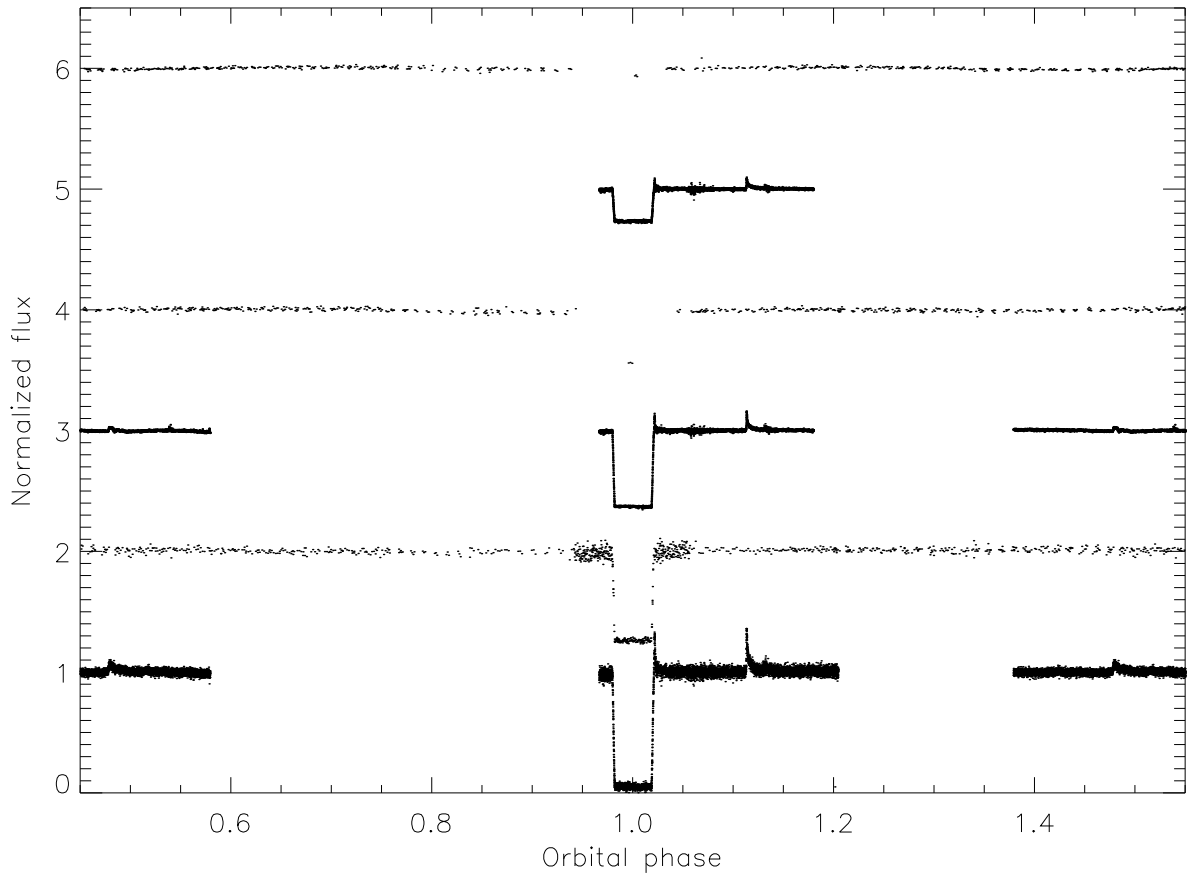}
\caption{The lightcurves of \rxj\ with the following filters (bottom to top);
 $u^{\prime}$, B,  $g^{\prime}$,  V, $r^{\prime}$  and
 I. The lightcurves have been normalized to the mean
out-of-eclipse values and offset by 1.0 for clarity. The data for the
$u^{\prime}$,  $g^{\prime}$ and  $r^{\prime}$ filters around primary eclipse
were  obtained on the
night 2002 May 17 and clearly show two flares. Data at other phases for the
$u^{\prime}$ and  $g^{\prime}$ bands were taken in poor seeing on the night 2003
November 13 and also shows two flares. \label{BVILCFig} }
\end{figure*}

 For a given inclination, $i$, a single eclipse will give an accurate
measurement of the radii of the stars relative to their separation, i.e.,
$r_{\rm WD} = R_{\rm WD} /a$ and $r_{\rm M} = R_{\rm M} /a$, where $a$ is the
separation of the stars and  $R_{\rm WD}$, $R_{\rm M}$ are the radii of the
white dwarf and the M-dwarf, respectively. Additionally, the depth of the
eclipse gives an accurate measurement of the luminosity ratio, $l_{\lambda}$
at the effective wavelength of the lightcurve $\lambda$. Equally good fits to
the eclipse can be found for a wide range of $i$ values. The  shape of the
M-dwarf is accurately determined by an equipotential surface in the Roche
potential, which is slightly non-spherical. This gives rise to an
ellipsoidal effect with an amplitude of about 2\,percent in the I~band
lightcurve.  The amplitude of the ellipsoidal effect varies strongly with
$r_{\rm M}$ and the predicted value of $r_{\rm M}$ increases as $i$ decreases,
so we have enough information in the $u^{\prime}$ band and I~band lightcurves
to measure $r_{\rm WD}$, $r_{\rm M}$ and $i$.

\begin{figure}
\includegraphics[width=0.49\textwidth]{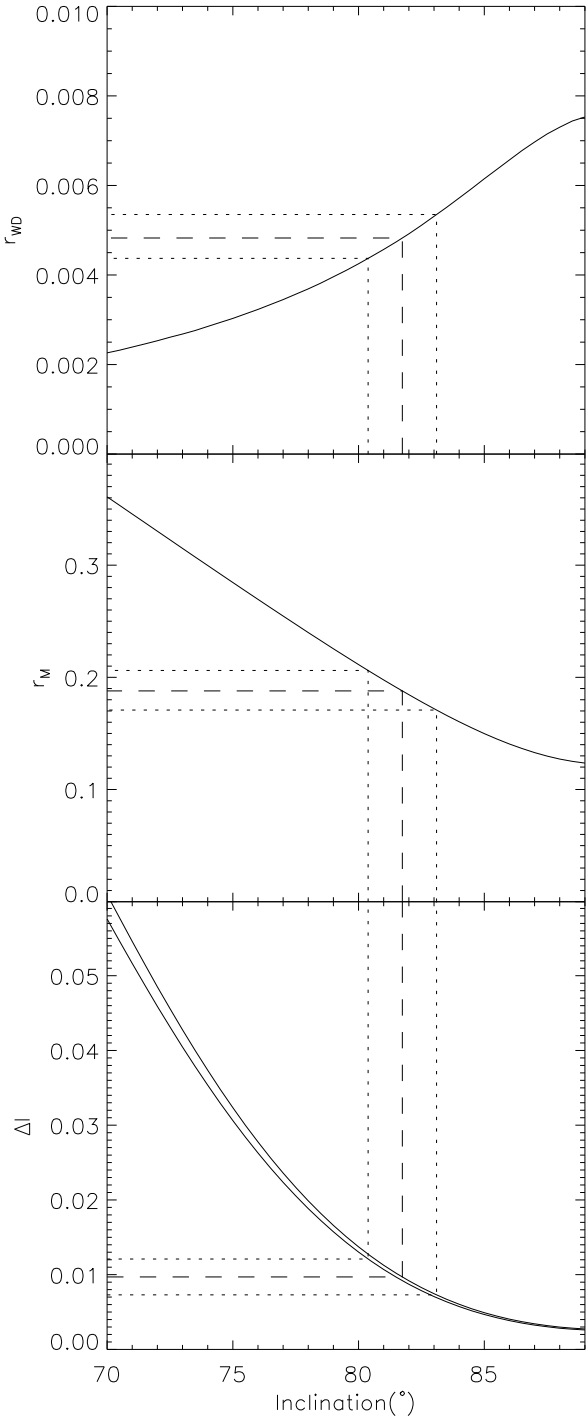}
\caption{Upper panel: The optimum value of $r_{\rm WD}$ as a function of
inclination from the least-squares fit to the $u^{\prime}$ band lightcurve;
middle panel: optimum value of $r_{\rm M}$; lower panel: the predicted
semi-amplitude of the I~band lightcurve, $\Delta I$ for two values of the limb
darkening coefficient.  The conversion of the observed value
of  $\Delta I$ to values of  $r_{\rm WD}$ and $r_{\rm M}$ and their
uncertainties are illustrated by
dashed and dotted lines, respectively.\label{LCParFig}} 
\end{figure}

 We first measured the values of $r_{\rm WD}$ and $r_{\rm M}$ as a function of
$i$ from the $u^{\prime}$ band lightcurve using the simple model of uniform
circular discs described in section \ref{EphemSect}. We included data obtain
on the night 2002 May 24 and 2002 May 25 in the fit after first normalising
the data using a straight line fit to the data outside of eclipse. The results
are plotted in Fig.~\ref{LCParFig}. We used the slightly more advanced
lightcurve model {\sc ebop} (Nelson \& Davis 1972, Popper \& Etzel 1981) to
investigate the effect of limb-darkening on the values of  $r_{\rm WD}$ and
$r_{\rm M}$ derived, we find that this radii change by less than 0.1\,percent
if limb-darkening is included in the model.  

 We used a simple model to calculate the semi-amplitude of the ellipsoidal
effect in the I~band lightcurve, $\Delta I$. The model creates a grid of points
over an equipotential surface in the Roche potential. The observed flux as a
function of phase is then calculated by numerical integration of the apparent
surface brightness over the points visible at each phase. The apparent surface
brightness over the star includes the effects of gravity darkening and
limb-darkening. We calculated the value of $\Delta I$ as a function of $i$ for
linear limb-darkening coefficients of  0.63  and 0.72, which covers the range
of values given in the tabulations of Claret (2000) for model stellar
atmospheres with T$_{\rm eff}$=3500\,K.  The gravity darkening exponent was
fixed at its standard value of 0.08. The effect of gravity darkening is to
make the lightcurve slightly fainter at phase 0 than phase 0.5. To calculate
$\Delta I$  we used the semi-amplitude calculated from the flux at phase 0.5
because the data at phase 0 are affected by the eclipse of the white dwarf,
although the difference is very small for the amplitudes we are dealing with
in this case.  We also ignore the effect of the light from the white dwarf
when we calculate $\Delta I$, so a small correction for this contribution to
the flux has to be made before comparing the predicted and observed values of
$\Delta I$. The predicted values of $\Delta I$ as a function of $i$  are shown
in Fig.~\ref{LCParFig}.

 The measurement $\Delta I$ is complicated by the asymmetry of the I~band
lightcurve, e.g., the minimum of the lightcurve does not occur at phase 0.5.
We assume that this is due to the presence of spots on the surface of the
M-dwarf.  This is not an unreasonable assumption given that this is a rapidly
rotating M-dwarf which is known to show flares (Fig.~8). To model this effect
of these spots on the lightcurve, we assumed that there are two spots  which
cause dips in the lightcurve described by the functions
$A_j(1+\cos[\phi -\phi_j])/2$, $j=1,2$, which are restricted to orbital phases
$\phi_j \pm \pi/2$. We modelled the ellipsoidal effect with the function
$-\Delta I \cos(2 \phi)$ and also included a component for the reflection
effect using  the function $-A_{\rm ref}\cos(\phi)$. We first formed the
average of the observations in phase bins 0.02 wide and calculated the
standard deviation of the mean of the data in each bin. We then subtracted a
constant to remove the contribution of the white dwarf to the measured flux
and renormalised the data by its mean. The value of the constant is determined
from the mean value of the observed flux during eclipse.

\begin{figure}
\includegraphics[width=0.49\textwidth]{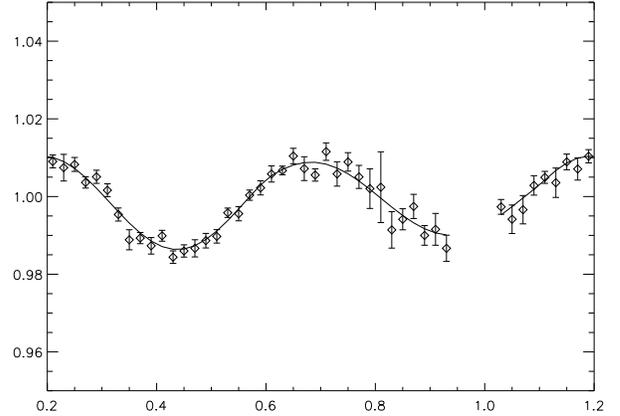}
\caption{The phase-binned I~band lightcurve of \rxj\ with a model fit with two
spot-like components, an ellipsoidal component and a reflection component.
\label{LCSpotFitFig}} 
\end{figure}

 We then performed a weighted least-squares fit of the sum of
the functions described above plus a constant to the phase-binned data
excluding the data in eclipse to find optimum values of the parameters $A_1$,
$A_2$ , $\phi_1$, $\phi_2$, $\Delta I$, $A_{\rm ref}$ and the constant. The
resulting fit and the phase-binned data are shown in Fig.~\ref{LCSpotFitFig}.
With so many parameters, there will clearly be many solutions which give a
good fit. In this case we are interested in finding the likely range of
$\Delta I$ values which are reasonable given the observed lightcurve. To find
the value of $\Delta I$ we attempted to fit the I~band lightcurve with the
dips due to the spots centered near phases 0.4 and 0.9 so as to remove as much
of the variability as possible using the spots alone. The parameters for this
fit are shown in Table~\ref{SpotFitTable}.

 The I-band data and the radial velocity data for the M-dwarf discussed in
Section~\ref{OrbitSect} were obtained simultaneously so the parameters for the
dark spots we infer by modelling distortions to the lightcurve and radial
velocity curve should be consistent. This is indeed the case for the phase of
the distortion of the proposed spot with $\phi_1 \approx 0.4$ which is also
responsible for the distortion to the radial velocity curve with $\phi_0 
\approx  0.2$ since $\phi_1 \approx \phi_0 + \rho/2$. The radial velocity
data near phase 0.8 are too sparse to confirm whether the value of $\phi_2 =
0.83$ for the second spot is reasonable. We can, in principle, check for
consistency between the amplitudes of the distortions, but this is a less
straightforward test (see Discussion below). We can at least point out that
the size of the distortions to the I-band lightcurves are about 2\,percent and
that the distortion to the radial velocity surve is about  2\,percent of the
equatorial rotational velocity of the M-dwarf (51\,\kms).

\subsection{Masses and radii of the stars}
\begin{figure}
\includegraphics[width=0.49\textwidth]{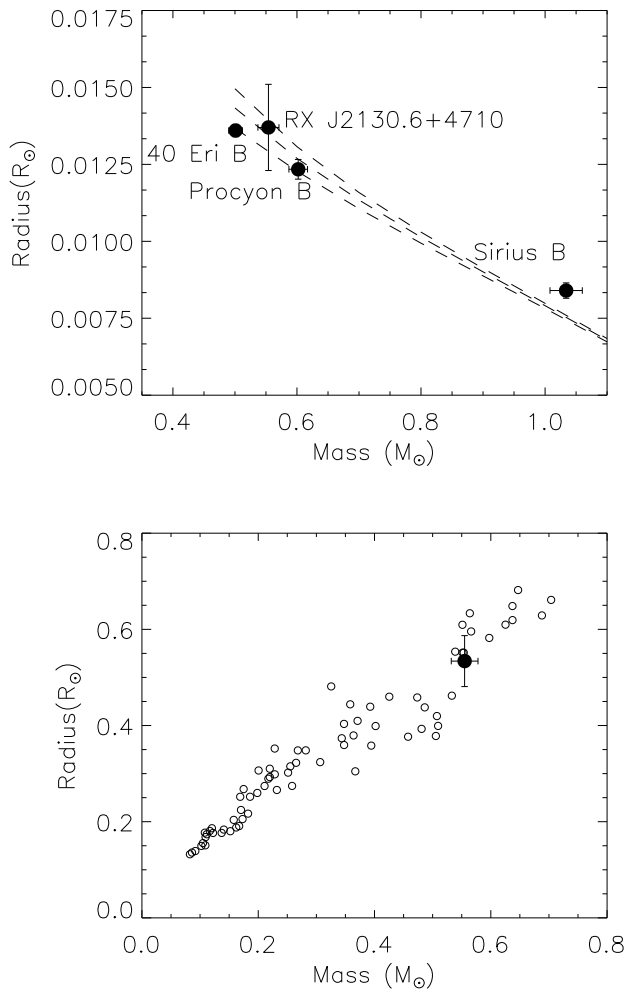}
\caption{Upper panel: The white dwarf component of \rxj\ in the mass-radius
plane compared to other
white dwarfs as labelled and the models of Benvenuto \& Althaus (1999)
for effective temperatures of  7\,500\,K, 17\,000\,K and 24\,790\,K.
Lower panel: The M-dwarf component of \rxj\ in the mass-radius plane (filled
circle) compared
to other
M-dwarfs (open circles) taken from Clemens et~al. (1998).
\label{MvRFig}} 
\end{figure}

 From the lower panel of Fig.~\ref{LCParFig} we find that the observed value
of $\Delta I = 0.0097\pm 0.0024$ corresponds to an inclination of $i =
(81.7\pm 1.4)^{\circ}$.  From the upper two panels we then find $r_{\rm
WD}=0.0048\pm 0.0005$ and $r_{\rm M}=0.188\pm 0.018$. We combine these values
with the combined spectroscopic orbit in Table~\ref{SpecOrbitTable} to derive
the masses and radii for the stars given in Table~\ref{MRTTable}. The
positions of the stars in the mass-radius plane is compared to other white
dwarfs and M-dwarfs in Fig.~\ref{MvRFig}.

\begin{table}
\caption{\label{SpotFitTable}Parameters of the least-squares fit to the I band
lightcurve.}
\begin{center}
\begin{tabular}{@{}lr}
\noalign{\smallskip}
Parameter & \multicolumn{1}{l}{Value}  \\
\noalign{\smallskip}
\hline
Constant      & $ 1.0103 \pm 0.00053 $ \\
$\Delta I$    & $ 0.0097 \pm 0.0024  $ \\
$A_{\rm ref}$ & $ 0.0010 \pm 0.0009  $ \\
$A_1        $ & $ 0.0175 \pm 0.0012  $ \\
$\phi_1     $ & $ 0.369  \pm 0.023   $ \\
$A_2        $ & $ 0.0143 \pm 0.0021  $ \\
$\phi_2     $ & $ 0.833  \pm 0.022   $ \\
$\chi^2$ &        40.86  \\
N        &            46 \\
\noalign{\smallskip}
\hline
\end{tabular}
\end{center}
\end{table}

\begin{table}
\caption{\label{MRTTable} The masses, radii and effective 
temperatures of the stars in \rxj.}
\begin{center}
\begin{tabular}{@{}lrr}
\noalign{\smallskip}
Parameter & \multicolumn{1}{l}{White dwarf} & \multicolumn{1}{l}{M-dwarf} \\
\noalign{\smallskip}
\hline
Mass (\Msolar) & $0.554 \pm 0.017 $ & $ 0.555 \pm 0.023 $ \\ 
Radius (\Rsolar)&$0.0137\pm 0.0014$&$0.534 \pm 0.053$ \\
$\Teff$ (K)& 18\,000 $\pm$ 1000 &  3200$\pm$100 $^{a}$ \\
\noalign{\smallskip}
\hline
\end{tabular}
\end{center}

$^a$ Based on our estimate of the spectral type and the
calibration of Leggett (1992).
\end{table}

\subsection{The effective temperature of the white dwarf}
 We compared the spectra of \rxj\ to a grid of synthetic spectra calculated 
from pure hydrogen model atmospheres to estimate the effective temperature,
$\Teff$, of the white dwarf. 

 These have been generated with the latest versions of the plane-parallel,
hydrostatic, non-local thermodynamic equilibrium (non-LTE) atmosphere and
spectral synthesis codes TLUSTY (version 200; Hubeny 1988,  Hubeny \& Lanz
2003) and SYNSPEC (version 48; Hubeny et~al. 1985). All calculations include a
full treatment of line blanketing and use a state-of-the-art model hydrogen
atom incorporating the 8 lowest energy levels and one super-level extending
from n=9 to n=80. During the calculation of the model structure, lines of the
Lyman and Balmer series are treated by means of an approximate Stark profile.
However, as we use the Balmer lines to determine the effective temperature and
surface gravity of the white dwarf, it is important that we employ the best
available line broadening data. Therefore, during the spectral synthesis step,
detailed profiles for the HI lines are calculated from the Stark broadening
tables of Lemke (1997). We calculated synthetic spectra for model atmospheres
over a  range $ 15\,000\,K < \Teff < 65\,000\,K$ in steps of 5000\,K and
surface gravities in the range $6.5 < \log g < 8.5$ in steps of 0.5 ($\log g$
in cgs units.).

 In order to compare the spectra of \rxj\ to the model spectra, we first had
to remove the contribution of the M-dwarf from the spectra.  The spectra were
normalized by a constant value determined from the mean flux around
4600\AA. We interpolated the spectra onto a uniform wavelength grid of 2760
pixels over the wavelength range 3780\AA\,--\,5105\AA. We formed the average
of spectra in groups taken at the same orbital phase to within 0.04 phase
units.  The phase ranges were chosen such that spectrum of \rxj\ taken at
mid-eclipse was in a group by itself. We calculated a scaling factor to apply
to the mid-eclipse spectrum prior to subtracting it  from each  of the 24
phase-binned spectra by measuring the ratio of fluxes either side of the band
head near 4954\AA\ and the depth of the absorption feature near 4226\AA. The
scaling factor calculated was typically 0.25 at  4226\AA\ and 0.36 at 4954\AA.
We multiplied the mid-eclipse spectrum by a  linear function of wavelength
fitted to these values prior to shifting it to account for the radial velocity
then subtracting it from the phase-binned spectra. 

 We convolved the synthetic spectra with a Gaussian function to account for
the resolution of the observed spectra and then interpolated the synthetic
spectra onto the same wavelength grid at the phase-binned spectra. We were
then able to compare the synthetic and observed spectra by calculating the
$\chi^2$ statistic for each combination of observed and synthetic spectrum. We
did this separately for each Balmer line after applying a linear normalization
calculated from the flux in the line wings to all the spectra. We excluded
from the calculation of $\chi^2$ a region $\pm 100$\kms\ wide centered on the
Balmer emission lines from the M-dwarf, the Ca\,II H and K emission lines and
an emission line near 3905.5\AA\ which may be due to Si\,I.

\begin{figure*}
\includegraphics[angle=90,width=0.99\textwidth]{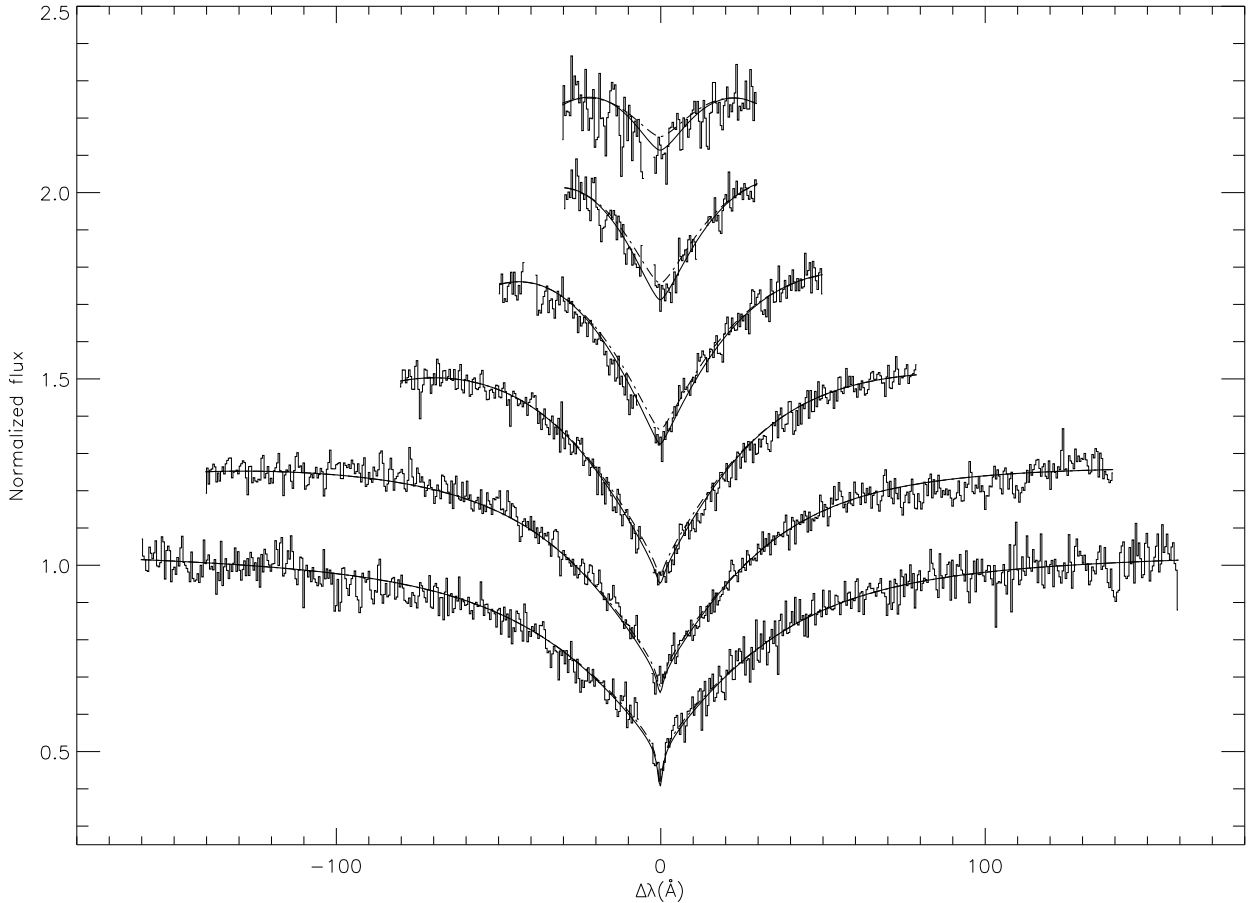}
\caption{A typical phase-binned spectrum of the white dwarf in  \rxj\ plotted
in sections offset by 0.25 around each Balmer line. The solid and dashed lines
are synthetic spectra from model atmospheres with $\Teff = 18\,000$K, $\log g =
7.84$  and $\Teff = 17\,000$K, $\log g = 7.99$, respectively.
\label{BalmerFitFig}} 
\end{figure*}

 For 20 of the phase-binned spectra the synthetic spectrum which gave the
lowest value of $\chi^2$ was calculated from model atmosphere with $\Teff =
20\,000\,K$, $\log g = 8.0$.  There is good agreement between the value of
$\log g$ derived from fitting these spectra and the value calculated from the
measured mass and radius, i.e., $\log g = 7.93 _{-0.09}^{+0.07}$. Therefore,
we created synthetic spectra corresponding to a range $ 15\,000\,K < \Teff <
20\,000\,K$ in steps of 1000\,K and surface gravities of $\log g = 7.93$, 7.84
and 8.00 by interpolation over the synthetic spectra calculated from the model
atmospheres. We calculated the $\chi^2$ statistic for these interpolated
spectra as before. We found that 17 of the spectra were best fit by model
spectra with  $\Teff = 18\,000$K, $\log g = 7.84$  or  $\Teff = 17\,000$K, $\log
g = 7.99$. These model spectra and a typical phase binned spectrum are shown
in Fig.~\ref{BalmerFitFig}.  The remaining 7 spectra were best fit by spectra
within one grid-step of these two models. The quality of fits for spectra
taken at phases from 0 to 0.5 judged by-eye and from the value of $\chi^2$  is
very good. The fits are not so good in the other half of the orbit if judged
by the value of  $\chi^2$ (reduced  $\chi^2$ values $\approx 1.5$). There is
no reason for this which is obvious from an inspection of the fits, but it is
probably due to a combination of inaccurate subtraction of the M-dwarf
spectrum and inaccuracies in the flux calibration.  The spectra in the phases
range  0 to 0.5 tend to be fit best by spectra with $\Teff = 18\,000$K.
Therefore, we adopt an effective temperature of $\Teff = 18\,000$K for the white dwarf in
\rxj\ and estimate that the uncertainty in this estimate is about 1000\,K.

\section{Discussion}

 We can estimate the distance and reddening to \rxj\ from its effective
temperature and our B and V photometry. From the tabulations of Bergeron,
Wesemael \& Beauchamp we estimate that the white dwarf in \rxj\ has an
intrinsic color (B$-$V)$_0$ = $0.00 \pm 0.02$ and an absolute V magnitude,
M$_{\rm V} =  11.0 \pm 0.1$. From the depths of the eclipses in the B and
V bands combined with the photometry in Table~\ref{MagTable} we can estimate
apparent magnitudes for the white dwarf of V=15.05$\pm$0.1,
(B$-$V)=0.2$\pm$0.14. From these values we estimate E(B$-$V)=0.2$\pm$0.15,
(m$-$M)=4.65$\pm$0.5 and a distance to \rxj of 85$\pm$20\,parsecs.

The distance to  HD\,204906 is known to be 82$\pm$5\,parsecs from its
Hipparcos parallax (Perryman 1987). It would be fortunate if \rxj\ could be
shown to be physically related to HD204906 because we could then measure 
accurate absolute magnitudes for the stars in this binary. Unfortunately, the
measured radial velocity for HD204906 is 34.8$\pm$1.4\,km\,s$^{-1}$ (Sandage
\& Fouts 1987), which is very different from the value of $\gamma_M =
15.3\pm0.7$ we have measured. This suggests that HD\,204906 and \rxj\ are not
physically related. One possibility that should be investigated is whether
HD\,204906 is itself a binary star.
 
 The M-dwarf in \rxj\ shows all the characteristics of a magnetically active
star, i.e., soft X-ray emission, Balmer emission lines, flares and spots.
We have attempted some crude modeling of the distortions to the radial
velocity and lightcurves to account for these distortions. In reality, these
spots are likely to be a complex pattern of dark regions with variations in
the local temperature and the emergent spectrum. This pattern of dark regions
can approximated using models with  circular spots, each with its own position
(latitude and longitude), physical size and temperature. Modeling the two
spots we have used to model the I-band lightcurve would therefore require
eight free parameters plus some model for how the emergent spectrum depends on
local temperature and viewing angle. This is not warranted given the quality
of our data. Further observations of \rxj\  with improved phase coverage and
resolution could be used to study the pattern of active regions on the surface
of the M-dwarf and the evolution of this pattern using techniques such as
Doppler imaging (Rice 2002) or monitoring of the lightcurve. Such studies
would benefit from the independent determinations of the rotational period of
the star (assuming it is the same as the orbital period) the inclination of
the star and the radius of the star presented here. This would, for example,
allow a comparison of the rotational period derived from a Fourier analysis of
the lightcurve excluding eclipses to the orbital period in order to look for
asynchronous rotation in the M-dwarf. Monitoring of the eclipse times will
then allow a further test of the mechanism proposed by Applegate (1992) to
explain period changes in binaries similar to \rxj\ such as V471~Tau. Indeed,
the orbital periods of V471~Tau and \rxj\ are very similar, so comparison of
the results for these two stars will provide observational evidence for the
way in which Applegate's mechanism depends on the mass of the magnetically
active star.

 In principle, a measurement of the depth of the secondary eclipse in \rxj\
would allow a much more accurate determination of the masses and radii of the
stars. Unfortunately, the predicted secondary eclipse depth is
7\,milli-magnitudes or less. This will be very difficult to measure given the
proximity of HD\,204906 to \rxj\ and the distortions to the lightcurve due to
spots and flares. A more straightforward way to improve the accuracy of the
masses and radii will be to obtain improved spectroscopy, both in terms of
resolution and signal-to-noise, and to obtain a lightcurve at near infrared
wavelengths where the measurment of the ellipsoidal effect will be less
affacted by star spots. High resolution spectroscopy at wavelengths of
3800\,--\,4000\AA\, may reveal sharp metal lines due to accretion of metals
onto the surface of the white dwarf from the stellar wind of the M-dwarf
(Zuckermann et~al., 2002). These lines, if present, would allow the
measurement of accurate radial velocites for the white dwarf and thus an
accurate measurement of its gravitational redshift.

 The spectral type of the M-dwarf in \rxj\ is estimated to be M3.5Ve or M4Ve
and its mass is $0.555 \pm 0.023$\Msolar. This combination of mass and
spectral type is not consistent with the calibaration of Leggett (1992) for
M-stars in the galactic disk. A typical M3.5V in the galactic disk has a mass
of about 0.2\,--\,0.3\,\Msolar. An M-type star in the galactic disk with a
mass of $ 0.55$\Msolar\ usually has a spectral type in the range M0V\,--\,M2V.
This discrepancy may be due to an inaccurate estimate of the spectral type for
the M-dwarf caused by using a limited spectral range with few features
sensitive to spectral type and comparing this spectrum to stars which are not
standard stars for the spectral class. A much more reliable estimate of the
spectral type of the M-dwarf in \rxj\ should be made using a spectrum taking
during the eclipse covering the TiO bands in the regions 6000\AA\,--\,9000\AA.
By contrast, the observed value of $(J-H)=0.63$ is negligibly affected by the
presence of the white dwarf and is typical for  M-type stars in the galactic
disk with a mass of $ 0.55$\Msolar.

 We have investigated the past and future evolution of \rxj\ using the analysis
described by Schreiber \& G\"{a}nsicke (2003).  The  cooling age, $t_{\rm
cool}$ of \rxj\ derived by interpolating the cooling tracks of Wood (1995) is
$\log(t_{\rm cool}/y) = 7.85$. This is the time since \rxj\ emerged from the
common-envelope phase, at which time the orbital period is estimated to have 
been 0.53\,--\,0.6\,d depending on the prescription used to model the angular
momentum loss due to a magentic stellar wind in this interval. The continued
loss of angular momentum will shrink the Roche lobe of the M-dwarf to the
point where mass transfer will start from the M-dwarf to the white dwarf
through the inner Lagrangian point. This will occur at an orbital period
$P_{\rm sd}=0.185\,d$. Our estimate of the time before Roche lobe overflow
occurs, $t_{\rm sd}$, also depends on the prescription for angular momentum
loss used and varies from $\log(t_{\rm sd}/y) =  8.1$ to $\log(t_{\rm sd}/y) =
9.6$. In either case, the time taken is less than a Hubble time so \rxj\ can
properly be described as a pre-cataclysmic variable star (pre-CV) in the sense
defined by Schreiber \& G\"{a}nsicke. The value of $P_{\rm sd}$ for \rxj\  is
well above the period gap for CVs and is second only to V471~Tau among the
systems compiled by Schreiber \& G\"{a}nsicke. This implies that \rxj\ is a
member of the group of the rare progenitors of long orbital period CVs.

\section*{Acknowledgments}
 PFLM and LMR were supported by a PPARC post-doctoral grant. The William
Herschel Telescope, Isaac Newton Telescope and Jocobus Kapteyn Telescope are
operated on the island of La Palma by the Isaac Newton Group in the Spanish
Observatorio del Roque de los Muchachos of the Instituto de Astrofisica de
Canarias. We would like to thank Romano Corradi for his prompt and informative
reply to our query regarding the timing accuracy of data from the ING.
This research has made use of the SIMBAD database, operated at CDS,
Strasbourg, France. We thank an anonymous referee for their comments and
suggestions regarding the modelling of distortions to the lightcurve and
 radial velocity curve.

\label{lastpage}
\end{document}